# R&D studies for the development of a compact transmitter able to mimic the acoustic signature of a UHE neutrino interaction

M. Ardid*, S. Adrián, M. Bou-Cabo, G. Larosa, J.A. Martínez-Mora, V. Espinosa, F. Camarena, M. Ferri

*Institut d'Investigació per a la Gestió Integrada de les Zones Costaneres, Universitat Politècnica de València, C/ Paranimf 1, 46730 Gandia, València, Spain.*
*Corresponding author. Tel.: +34 962849314; Fax: +34 962849309
E-mail address: mardid@fis.upv.es

## ABSTRACT

Calibration of acoustic neutrino telescopes with neutrino-like signals is an essential to evaluate the feasibility of the technique and to know the efficiency of the detectors. However, it is not straightforward to have acoustic transmitters that, on one hand, are able to mimic the signature of a UHE neutrino interaction, that is, a bipolar acoustic pulse with the 'pancake' directivity, and, on the other hand, fulfill practical issues such as ease of deployment and operation. This is a non-trivial problem since it requires directive transducer with cylindrical symmetry for a broadband frequency range. Classical solutions using linear arrays of acoustic transducers result in long arrays with many elements, which increase the cost and the complexity for deployment and operation. In this paper we present the extension of our previous R&D studies using the parametric acoustic source technique by dealing with the cylindrical symmetry, and demonstrating that it is possible to use this technique for having a compact solution that could be much more easily included in neutrino telescope infrastructures or used in specific sea campaigns for calibration.

1. Introduction

Acoustic detection of Ultra High Energy (UHE) neutrinos is a promising technique which could open a window for the understanding of the most energetic processes in the Universe. However, the field is still in the phase of studying the feasibility with different acoustic systems that are being evaluated [1-4]. In this context the development of acoustic transmitters able to monitor the sensitivity of the different acoustic sensors, to train and tune the detector, and to test the reliability of these system is of great importance [5]. Transmitters able to reproduce the acoustic signature of a UHE neutrino could perform most of these tasks for undersea neutrino telescopes, but due to the nature of the acoustic pulse (i. e., a very directive 'pancake' short bipolar pulse), it is not straightforward to design such a system, being able to reproduce the time and directivity patterns at the same time, especially if practical issues such as ease of deployment and operation are considered. The use of a linear array of omnidirectional emitters has been proposed for this task [6]. A different approach was considered in [7], where a first evaluation of the non-linear parametric acoustic sources effect [8-9] was done using planar transducers. The main advantage of this approach is that the transmitter is more compact, and therefore the difficulties for the deployment and operation in deep sea are smaller. In this paper, we extent our previous studies to the case of sources with cylindrical symmetry. Although the theory behind is the same, to achieve the cylindrical symmetry is, in practice, a challenge due to the facts of a larger geometric attenuation, scarcity of transducers available for it, and the lack of experiences and literature in the use of the acoustic parametric sources effect for transient signals using cylindrical transducers. In next sections we describe the experimental setup, present the different studies performed to evaluate the technique and analyze the results and, finally, summarise the work with the conclusions and discuss the future work.

2. Experimental Setup

The experimental setup follows the classical scheme of confronted emitting and receiving transducers in a water tank (1.10 x 0.85 x 0.80 m³). The emitter used was a Free Flooded Ring hydrophone (Sensortech SX83). This is a transducer with cylindrical symmetry which has the main resonance frequency at 10 kHz. It has a second resonance peak at 380 kHz. This frequency was chosen for our studies. The transducer was fed with a signal using a function generator (National Instruments PCI-5412) and a linear 55 dB RF amplifier (ENI 1040L). To measure the acoustic waveforms a spherical omni-directional transducer (National Instruments ITC-1042) connected to a 24 dB gain preamplifier (Reson CCA 1000) and a digitizer card (National Instruments PCI-5102) were used. With this configuration the receiver presents an almost flat frequency response below 100 kHz with a sensitivity of about -180 dB, whereas it is 38 dB less sensitive at 380 kHz. This election was made to be much more sensitive to the bipolar pulse (10-50 kHz) than the primary beam (380 kHz). A three-axis micro-positioning system was used to move the receiver in three orthogonal directions with a nominal accuracy of 10 μm. All the signal generation and acquisition process is based on a National Instruments PXI-Technology controller NI8176, which also control the micro-positioning system. With this it was possible to make the different sets of measurements as a function of the position in an automatic way.

3. Studies to evaluate the technique and results.

In this section we present a study to reproduce the neutrino acoustic signal from a cylindrical transducer using the parametric effect. Although this lab study is still far from the real conditions in a neutrino telescope it can be very useful to evaluate the feasibility of the technique and set the basis for the final design.

3.1. Studies of the shape of the parametric signal

Based on the knowledge about parametric generation with transient signals [5,8] to determine the signal for the emission we have used the following expression:

$$p(x,t) = \left(1 + \frac{B}{2A}\right) \frac{P^2 S}{16 \pi \rho c^4 \alpha x} \frac{\partial^2}{\partial t^2}\left[f\left(t - \frac{x}{c}\right)\right]^2$$

where $P$ is the pressure amplitude of the primary beam pulse (the modulation at 380 kHz, in our case), $S$ the surface area of the transducer, $f(t-x/c)$ is the envelop function of the signal, which is modulated at the primary beam frequency, $x$ is the distance, $t$ is the time, $B/A$ the nonlinear parameter of the medium, $\rho$ is the density, $c$ the sound speed and $\alpha$ is the absorption coefficient. The function $p(x,t)$ is the expected parametric signal (pressure as a function of the distance from the transducer $x$, and the time $t$). With the aim of demonstrate the capacity of controlling the parametric signal generated in the medium, we have performed different measurements using as default the bipolar signal given by the first time derivative of a Gaussian function with a sigma of 5 μs. The envelope function f(t-x/c) was calculated by integrating twice the expression. The shape of this signal was varied either using a different sigma: 2.5, 5, 10 or 20 μs, or by separating the positive and negative parts of the bipolar pulse by adding in the middle of the signal a constant amplitude signal with lengths: 0, 5, 10, 20 or 50 μs. Figure 1 shows some of the signals used for emission. The received signals were, in general, a mix of the primary beam at 380 kHz (main component) and the secondary transient signal produced by parametric effect. In order to better distinguish these components, a band pass filter ([250 450] kHz) was applied for the primary beam, whereas a band pass [5 100] kHz filter was used for the secondary parametric beam. Figure 2 shows some examples of received signals (dark blue), and primary (light blue) and secondary beams (red). An interesting parameter for this study is the time separation between the maximum and minimum of the signals, δt. As expected, Δt increases with the sigma used, from 7-8 µs (for 2.5 and 5.0 µs sigma) to 10-11 µs (for 10 and 20 µs sigma). With respect to the constant amplitude added in the middle of the signal, for small lengths it results in a moderate increase of δt, but no clear separation between positive and negative parts is observed. However if the length is very large (50 µs), this separation is clearly observed, as shown in Figure 2. Therefore, from this study we can conclude that, as expected, the shape of bipolar signal is, approximately, the second time derivate of the primary signal envelope, and we have some different possibilities to control the parametric signal we would like to generate.

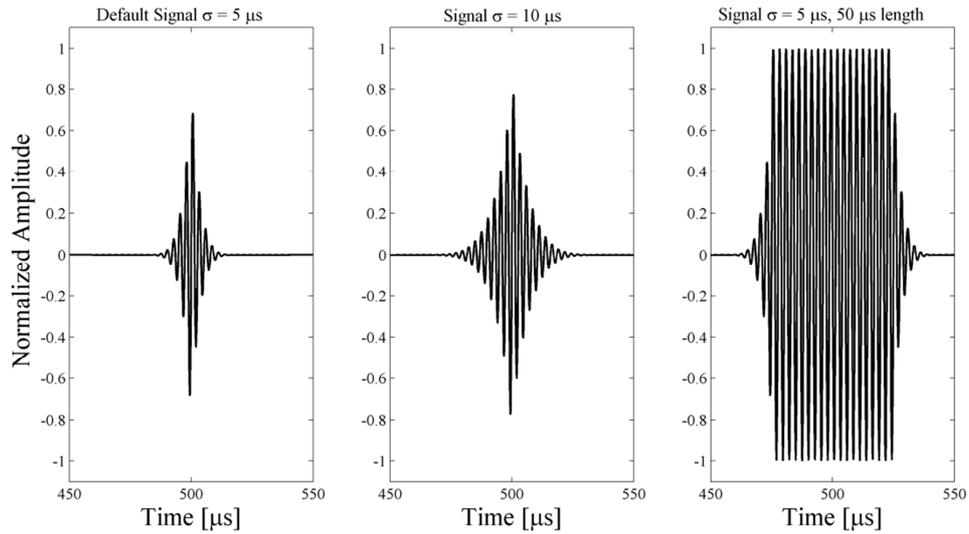

Fig. 1. Examples of different signals used for emission to study the shape dependences of the secondar parametric bipolar pulse. Left: the default signal with a 5 ms sigma. Centre: signal with a 10 ms sigma. Right: signal with a 5 ms sigma but with 50 ms length constant amplitude in the middle.

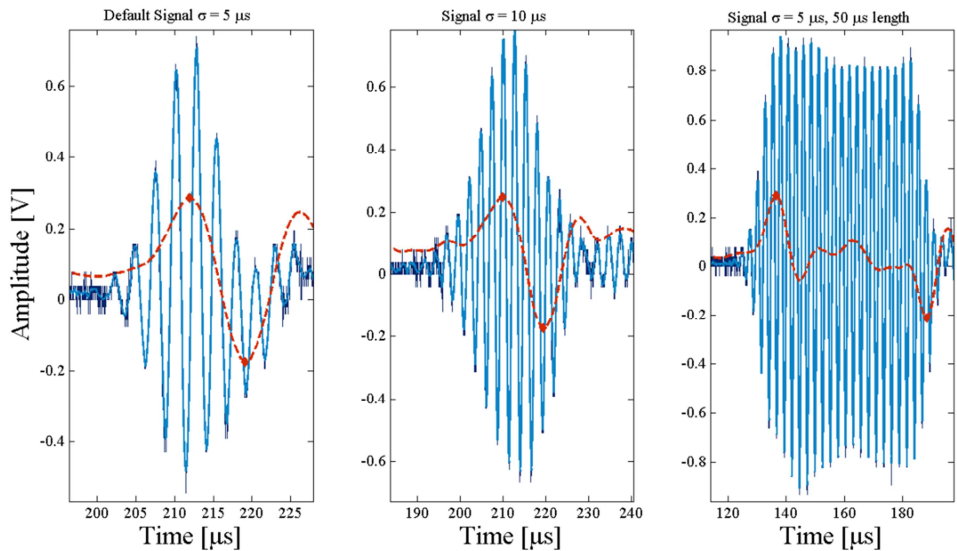

Fig. 2. Received signals (thin dark blue), primary beam (thick light blue) and parametric secondary beam, i.e. bipolar signal (red dashed line) for the emitted signals shown in Fig. 1. The bipolar signal (red dashed line) has been multiplied by a factor of 3 for a better visual comparison of the different signals. The maximum and minimum of the bipolar signal used to calculate δt and δV are highlighted. Notice that for the last signal, the constant amplitude between the rise and fall of the primary beam results in a separation of the positive and negative parts of the bipolar pulse. (For interpretation of the references to color in this figure legend, the reader is referred to the web version of this article.)

### 3.2. Studies of the parametric signal generation with the intensity of the primary beam.

We have evaluated the parametric generation as a function of the tension amplitude in the transducer by means of comparing the amplitude (peak-to-peak, $\delta V$) of the primary and secondary beams as a function of this parameter. From theory we expect a linear relationship for the primary beam and a nonlinear behaviour of the secondary beam (proportional to the square of the amplitude of the input signal). Figure 3 shows the results of this study using the default signal, that is, the short signal with sigma 5 µs. Fitting the data with potential functions, we observe that the exponent for the secondary beam is twice the exponent of the primary beam, which agrees with theory. However, the exponent for the primary beam is slightly below one. The reason for this deviation is probably due to saturation effects in the transducer part.

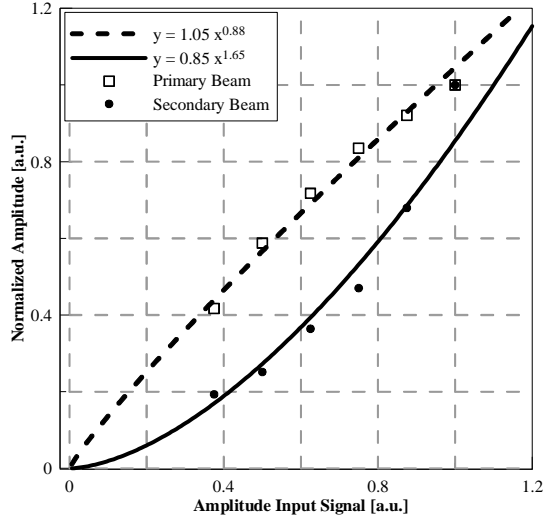 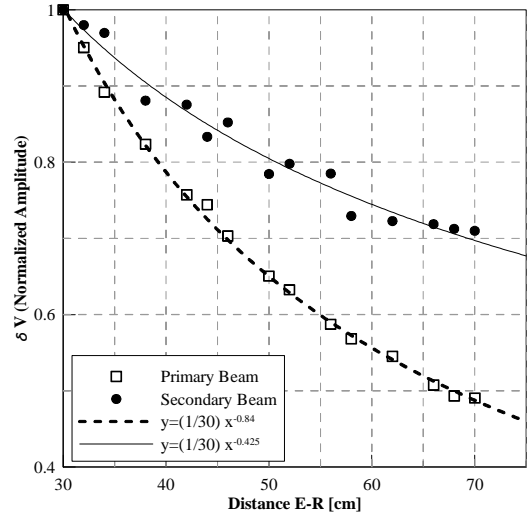

Fig. 3. Normalized amplitude of the received (primary beam) and bipolar (secondary beam) signal as a function of the amplitude input tension in the transducer. Potential functions have been fitted to the data and are shown in the plot.

Fig. 4. Distance dependence of the amplitudes of the primary and secondary parametric beams. Potential functions have been fitted to the data and are shown in the plot.

### 3.3. Studies of the parametric signal as a function of the distance from the transducer

The dependence of the parametric signal with the distance from the transducer has been studied by looking at the $\delta V$ amplitude of the primary and secondary beam as a function of the distance (at 0 degrees, that is, emitter and receiver aligned). The results of this study using the default signal can be observed in Figure 4. Potential functions have been fitted to the data with good agreement. For the primary beam, we expected an exponent of -1 for an omni-directional (spherical) transducer, and -0.5 for an ideal non-diverging cylindrical transducer. Since our transducer is cylindrical, but it is clearly diverging, the value of -0.86 seems reasonable. For the secondary beam we expected a significantly smaller exponent since parametric generation is being produced in the medium, at least for small distances. At distances of about 1 m, the amplitude of the bipolar pulse is two orders of magnitude lower than the amplitude of the primary beam. However, considering the attenuation and the higher absorption at high frequency, at distances larger than a few hundred meters the bipolar pulse is dominant. Therefore, for the application in underwater neutrino telescopes, it can be considered a 'clean' technique.

### 3.4. Directivity studies

The main advantage of using the parametric technique is the possibility of having broadband low-frequency directive beams, which is an essential aspect to have bipolar signals with the 'pancake' directivity. Here, we have measured and compared the directivity obtained for the primary and secondary beams using this cylindrical symmetry transducer. The results for the default signal are shown in Figure 5. It can be observed that, despite the differences in the spectral content, both, primary and secondary beams have similar pattern directivities.

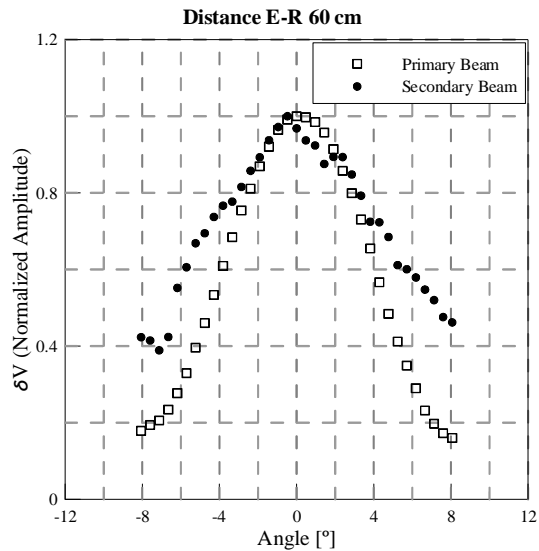

Fig. 5. Directivity pattern for the primary beam and the secondary bipolar signal measured in the tank with 60 cm distance between the emitter transducer and the receiver.

### 4. Conclusions and future steps

The laboratory tests performed to understand the control of the shape of the bipolar signal (secondary parametric beam), the studies to know the amplitude as functions of the tension in the transducer, and of the distance to it, as well as, the measurement of the directivity pattern show that the acoustic parametric sources effect is a promising tool that could be used to generate neutrino like signals with good directivity using a cylindrical transducer (or a compact array with a few of them). Moreover, with this kind of transducers, a linear array of 3 transducers with 20 cm separation from each other could reproduce the 1° 'pancake' directivity. Specific electronics for the feeding and operation of this transmitter will be developed using the concepts presented in [10-12]. The final compact transmitter will be tested and used in the summer 2011 with the deep sea positioned system of acoustic sensors AMADEUS [1,13], which will be useful to test the transmitter, and to perform some calibrations in the AMADEUS system.
For the next future, since the parametric effect is weak, it would be desirable to improve the efficiency of the transducers for the parametric bipolar pulse generation, and, for this, a study of different transducers options is also foreseen.

**Acknowledgements**

This work has been supported by the Ministerio de Ciencia e Innovación (Spain Government), project references FPA2007-63729, FPA2009-13983-C02-02, ACI2009-1067, Consolider-Ingenio Multidark (CSD2009-00064). It has also being funded by Generalitat Valenciana, Prometeo/2009/26.